\documentclass{article}

\usepackage[preprint,nonatbib]{neurips_2022}

\usepackage[utf8]{inputenc} 
\usepackage[T1]{fontenc}    
\usepackage{url}            
\usepackage{booktabs}       
\usepackage{nicefrac}       
\usepackage{microtype}      
\usepackage{xcolor}         
\usepackage{times}
\usepackage{soul}
\usepackage{tabularx}
\usepackage{bbm}
\usepackage{dsfont}
\usepackage{graphicx}
\usepackage{multicol}
\usepackage{makecell}
\usepackage{amsmath,amsthm,bm,multirow}
\usepackage{subfigure}
\usepackage{pifont}
\usepackage{float}
\usepackage[normalem]{ulem}

\makeatletter
\newcommand*{\rom}[1]{\expandafter\@slowromancap\romannumeral #1@}
\makeatother
\usepackage{enumitem}
\usepackage{comment}

\usepackage{algorithmic}
\usepackage{algorithm}
\usepackage{diagbox}
\usepackage{textcomp}

\newcommand{\squishlist}{
    \begin{list}{$\bullet$}
        { \setlength{\itemsep}{0pt}      \setlength{\parsep}{0pt}
            \setlength{\topsep}{0.5pt}       \setlength{\partopsep}{0pt}
            \setlength{\listparindent}{-2pt}
            \setlength{\itemindent}{-5pt}
            \setlength{\leftmargin}{0.5em} \setlength{\labelwidth}{0em}
            \setlength{\labelsep}{0.2em} } }
    
\newcommand{\squishend}{
\end{list}  }

\usepackage[capitalize,noabbrev]{cleveref}
\usepackage[textsize=tiny]{todonotes}
\usepackage{wrapfig}

\theoremstyle{plain}

\theoremstyle{definition}

\theoremstyle{remark}

\usepackage{enumitem}

\title{S4: a High-sparsity, High-performance AI Accelerator}

\author{%
  Ian En-Hsu Yen \\
  Moffett AI \\
  Los Altos, CA 94022 \\
  \texttt{ian.yan@moffett.ai} \\
   \And
   Zhibin Xiao \\
   Moffett AI \\
   Los Altos, CA 94022 \\
   \texttt{zb.xiao@moffett.ai} \\
   \AND
   Dongkuan Xu \\
   The Pennsylvania State University \\
   North Carolina State University \\
   State College, PA 16802 \\
   \texttt{dongkuanx@gmail.com} \\
}

\begin{document}

\newcommand{\sysname}{\texttt{AutoDistil} }
\newcommand{\sysnameOnlyAuto}{{\footnotesize{ \texttt{AutoDistil}}}}
\newcommand{\cmark}{\ding{51}}%
\newcommand{\xmark}{\ding{55}}%

\maketitle

\begin{abstract}

Exploiting sparsity underlying neural networks has become one of the most potential methodologies to reduce the memory footprint, I/O cost, and computation workloads during inference. And the degree of sparsity one can exploit has become higher as larger model sizes have been considered along with the trend of pre-training giant models.
On the other hand, compared with quantization that has been a widely supported option, acceleration through high-degree sparsity is not supported in most computing platforms.
In this work, we introduce the first commercial hardware platform supporting high-degree sparsity acceleration up to 32 times --- S4. Combined with state-of-the-art sparse pruning techniques, we demonstrate several-times practical inference speedup on S4 over mainstream inference platforms such as Nvidia T4. We also show that in practice a sparse model of larger size can achieve both higher accuracy and higher throughput on S4 than a dense model of smaller size. 

\end{abstract}
\section{Introduction}
Deep neural network models have significantly improved the performance of various natural language processing (NLP)~\cite{devlin2019bert,NEURIPS2020_1457c0d6,xu2022autodistil} and computer vision (CV)~\cite{dosovitskiy2020image,liu2021swin} tasks in the recent years. While effective and prevalent, these models are usually prohibitively large. An emerging subfield has studied the redundancy in deep neural network models~\cite{zhu2017prune,gale2019state}, exploiting the sparsity of deep neural network models and finding sparse equivalent sub-network ~\cite{frankle2019}.
Moreover, along with the trend of pre-training giant models, such as BERT~\cite{devlin2019bert}, ViT~\cite{dosovitskiy2020image}, and GPT-3~\cite{NEURIPS2020_1457c0d6}, larger model sizes have been considered, which yields sparse sub-network of higher degree of sparsity. 

However, in contrast to quantization~\cite{han2015deep,kim2021bert} that has been widely adopted as a standard option for acceleration, most computing platforms do not support acceleration through high degrees of sparsity. Only the newly released Nvidia A100 starts to support sparse tensor operations as an acceleration option (up to 2x). As a result, most existing sparsity research can hardly lead to practical speedup on high-performance computing platforms.

To fill this gap, we introduce S4, the first commercial hardware platform that supports high-degree sparsity acceleration up to 32 times. S4 is an inference platform for datacenter of similar hardware parameters to Nvidia T4, but with additional high-degree-sparsity support. Combined with state-of-the-art sparse pruning techniques, we demonstrate several-times practical inference speedup on S4 over mainstream inference platform Nvidia T4. We also show that in practice a sparse model of larger size can achieve both higher accuracy and higher throughput on S4 than a dense model of smaller size. 

    
    
    

\section{S4 Platform}

The architecture of S4 can be summarized as follows:

\begin{figure*}[!t]
\vspace{-0cm}
\begin{center}
\centerline{\includegraphics[width=0.95\textwidth,height=0.7\textwidth]{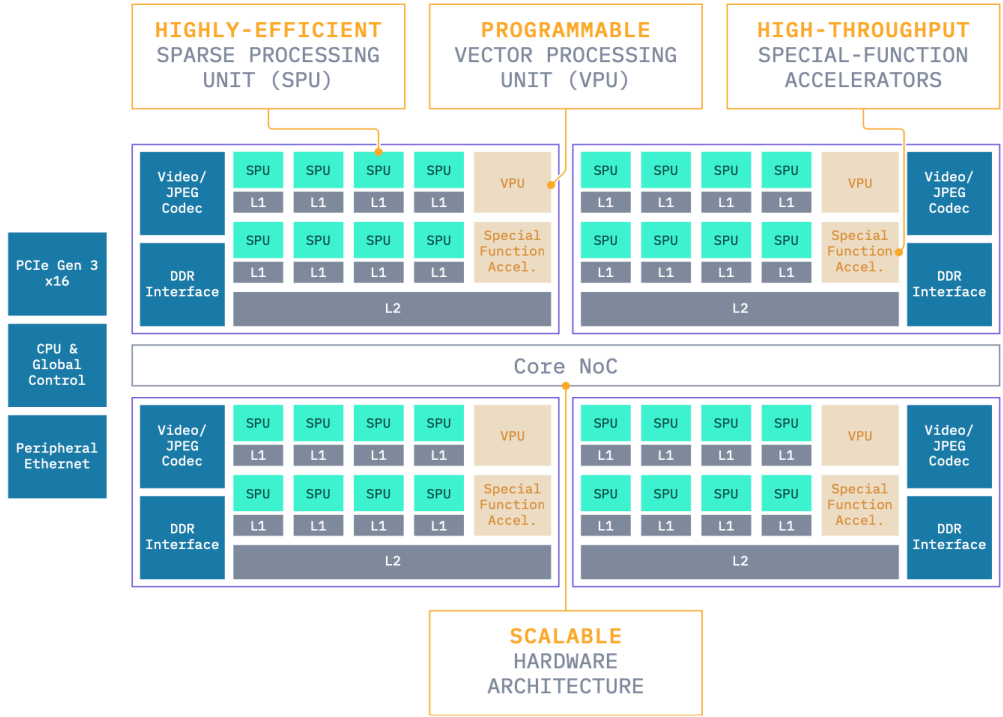}}
\caption{Architecture overview of Antoum processor: \textcolor{blue}{(i)} The sparse processing units (SPU) can support up to 32x tensor sparsity with linear speedup. \textcolor{blue}{(ii)} The customized activation engines directly support complex activation functions such as GELU, and basic mathematic operators such as exponential, log, reciprocal operators. \textcolor{blue}{(iii)} The sparse processing units natively support convolution and matrix multiplication operations with fused operations such as bias addition, elementwise operations, quantization, and certain activation functions. \textcolor{blue}{(iv)} Antoum moves the computation units directly adjacent to large capacity and large bandwidth memory banks.}
\label{fig:S4}
\end{center}
\vspace{-0cm}
\end{figure*}

\begin{itemize}
    \item High-rate sparse tensor kernel. S4 card is the first AI inference accelerator card that supports high-rate (up to 32x sparsity) sparse tensor operations.
    \item High-Performance Multimedia Processing Capability. The S4 card integrates dedicated video codec engines and JPEG decoder engines. The four video decoder engines and one video encoding engine can handle multi-channel video streams (up to 4K) and easily integrate scalable deep learning into video processing. 
    \item Scalability. The S4 card forms a sparse processing subsystem through a custom sparse processing unit and other auxiliary acceleration units, including dedicated video codec and JPEG decoder engines, embedding lookup units, memory reshape engine, and vector processors. Four sparse processing subsystems form a complete chip through a high-bandwidth, on-chip ring interconnection network. 
    \item The S4 hardware is supported through \emph{SparseRT} development toolkit, which supports existing AI programming frameworks such as Tensorflow, PyTorch, ONNX and MXNet.
\end{itemize}

Built to enhance the efficiency of AI inference in datacenters, S4 provides a (sparse) equivalent computation power of 944 TOPS in INT8 and 472 TFLOPS in BF16, and has 20GB LPDDR4 memory with up to 72 GB memory bandwidth in a low 70 Watt power envelope.
The combined effect of Moffett’s original sparsity algorithm and Antoum chip architecture has greatly increased the computation speed of S4, thus reducing the total cost of ownership (TCO). The Moffett Antoum architecture is shown in Figure~\ref{fig:S4}. The hardware and software are tightly engineered to create a highly efficient AI system-on-chip (SoC) processor platform. The combination of the sparse processor units (SPU) (for native sparse convolution and matrix multiplication) and the heterogenous unique function accelerators, provides maximum efficiency for various AI inference workloads and maximal value for all users. For example, the integrated Vector Processor Unit (VPU) can provide flexible programmability to keep up with the fast evolution of AI models. The on-chip video codec supports 64-way 1080p video decoding at 30 FPS. The JPEG decoder supports up to 2320 FPS 1080p image decoding, which provides a complete end-to-end solution for video and image inference workload.
\section{Sparse Acceleration on S4}

\begin{figure*}[!t]
\vspace{-0cm}
\centering
  \subfigure[ResNet50.]{\label{fig:perf_vs_sparsity-1}
  \includegraphics[width=0.475\textwidth,height=0.33\textwidth]{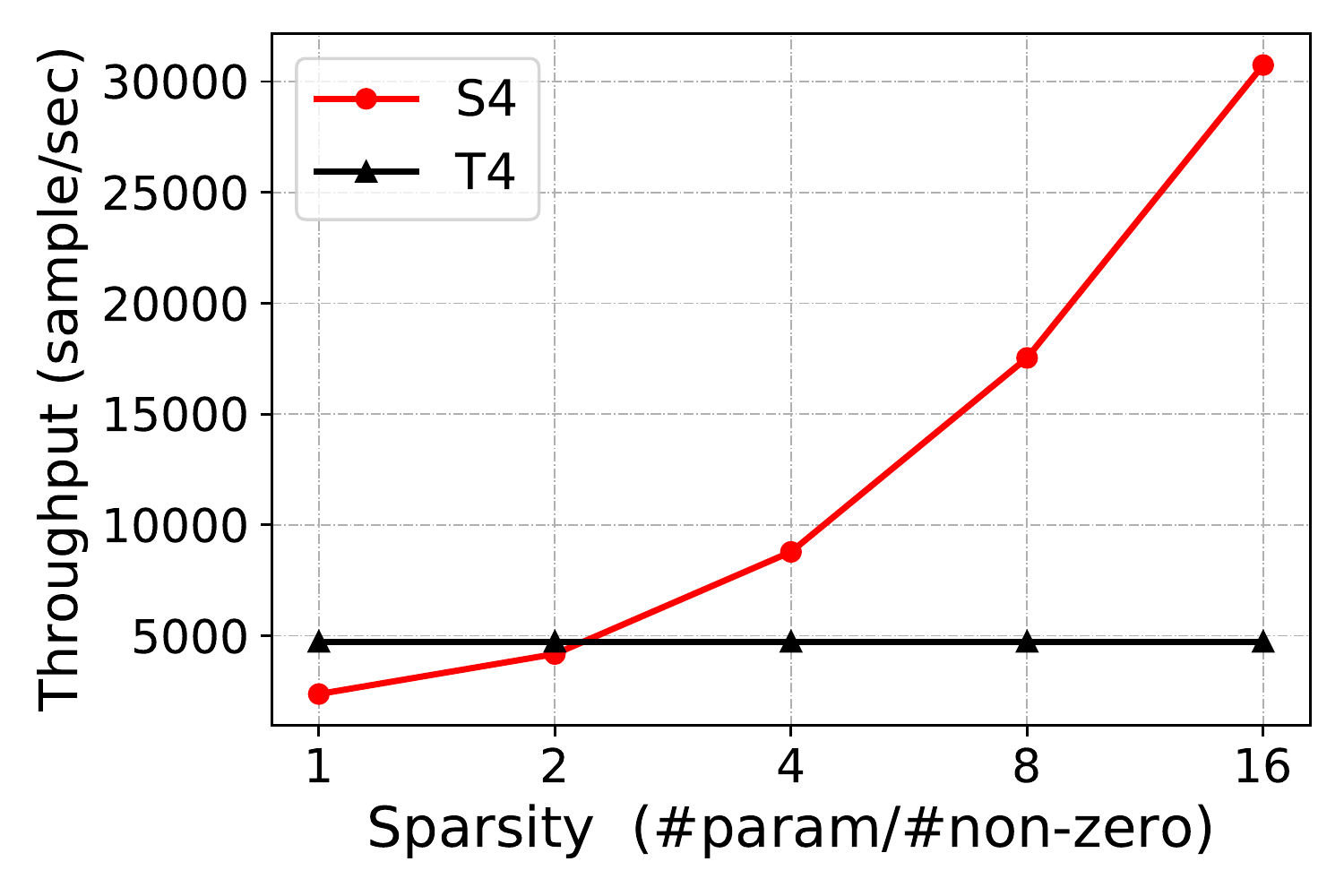}
  }\vspace{0cm}
  \subfigure[BERT.]{\label{fig:perf_vs_sparsity-2}
  \includegraphics[width=0.475\textwidth,height=0.33\textwidth]{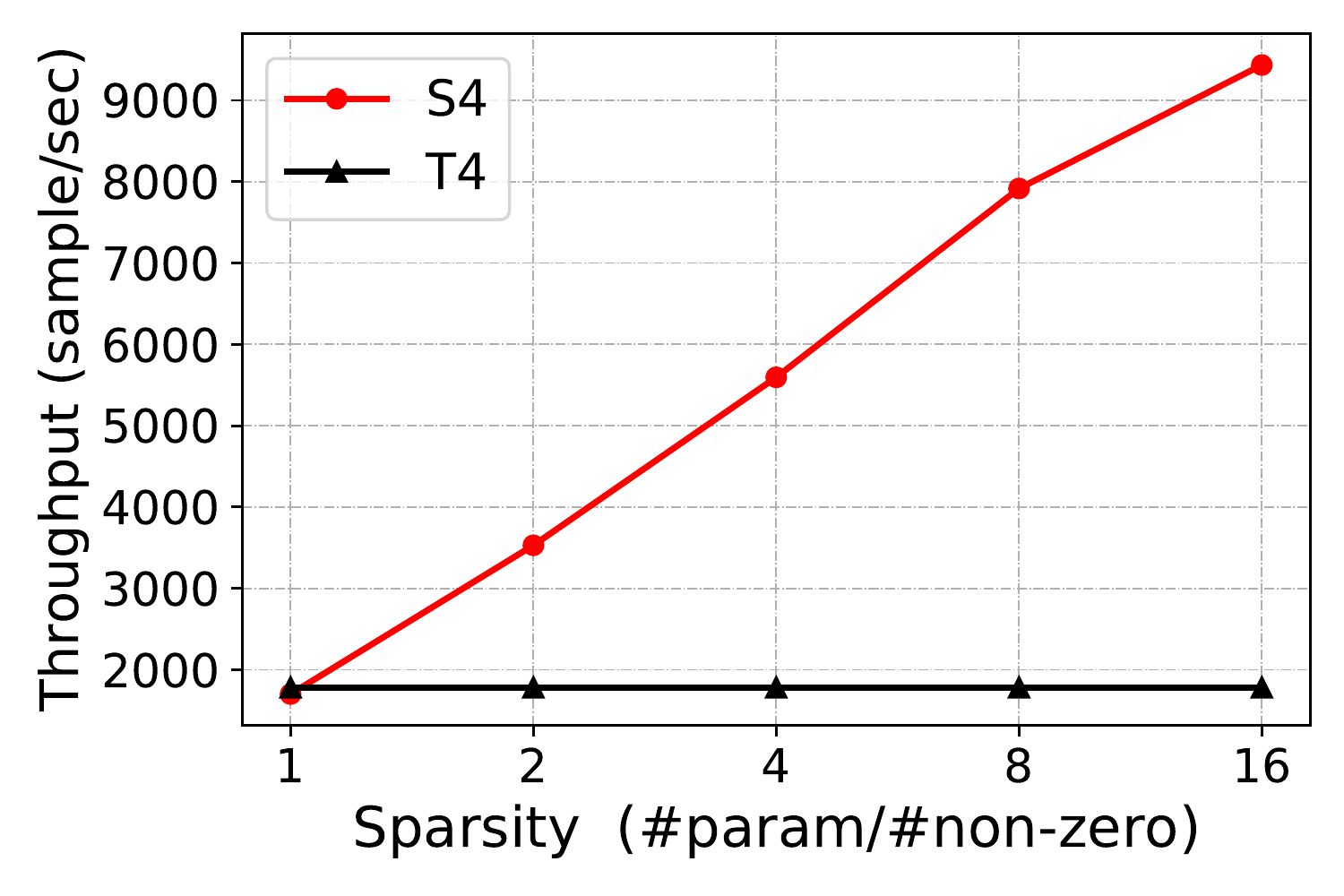}
  }
\caption{Speedup (throughput) achieved on Moffett S4 at different levels of sparsity, and a reference throughputs of Nvidia T4 from its official website \cite{nv_product_benchmark_link}. } 
\label{fig:perf_vs_sparsity}
\end{figure*}

\begin{figure*}[!t]
\centering
  \subfigure[ResNet50 and ResNet152 on ImageNet.]{\label{fig:acc_vs_sparsity-1}
  \includegraphics[width=0.475\textwidth,height=0.33\textwidth]{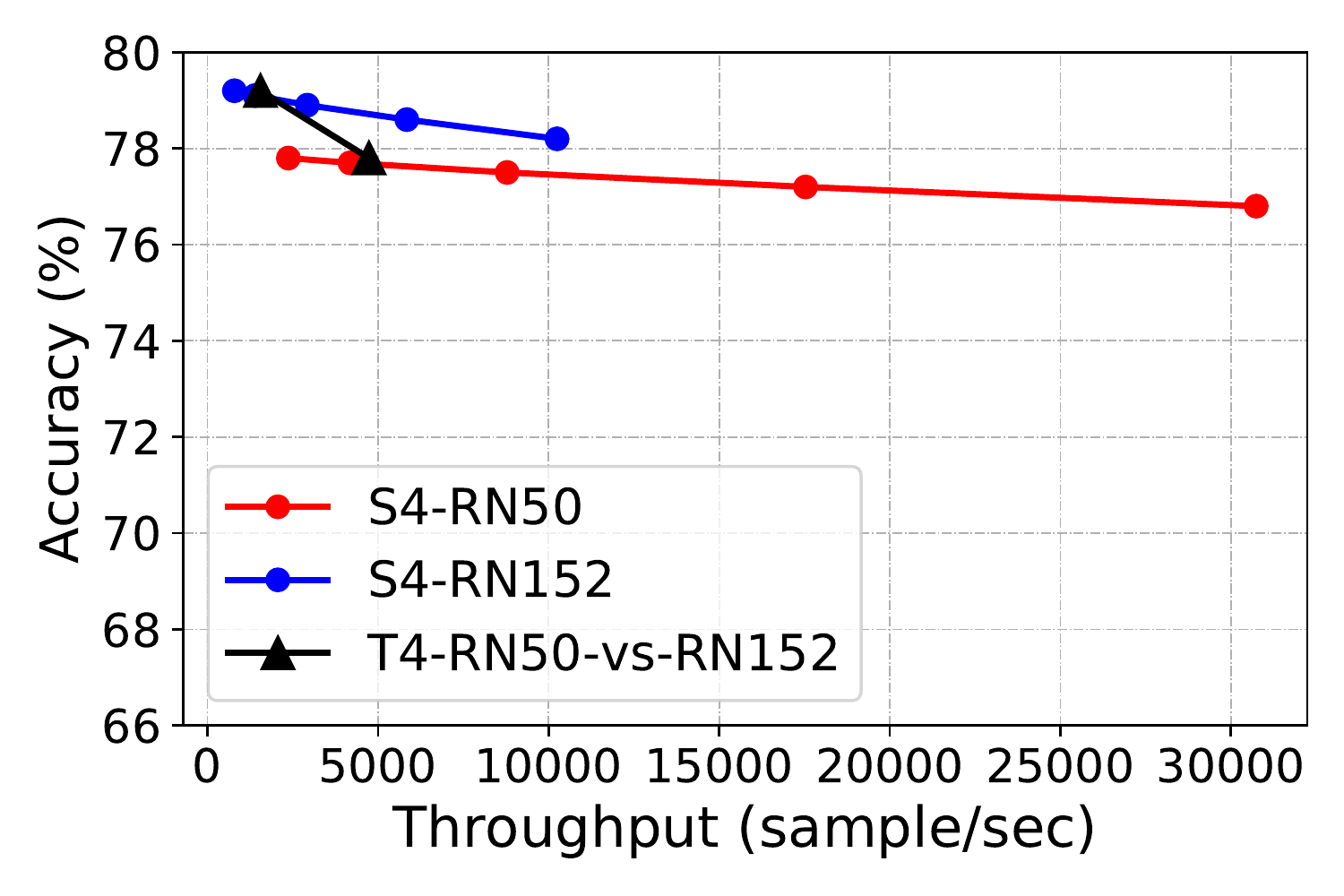}
  }\vspace{0cm}
  \subfigure[BERT-Base and BERT-Large on SST2.]{\label{fig:acc_vs_sparsity-2}
  \includegraphics[width=0.475\textwidth,height=0.33\textwidth]{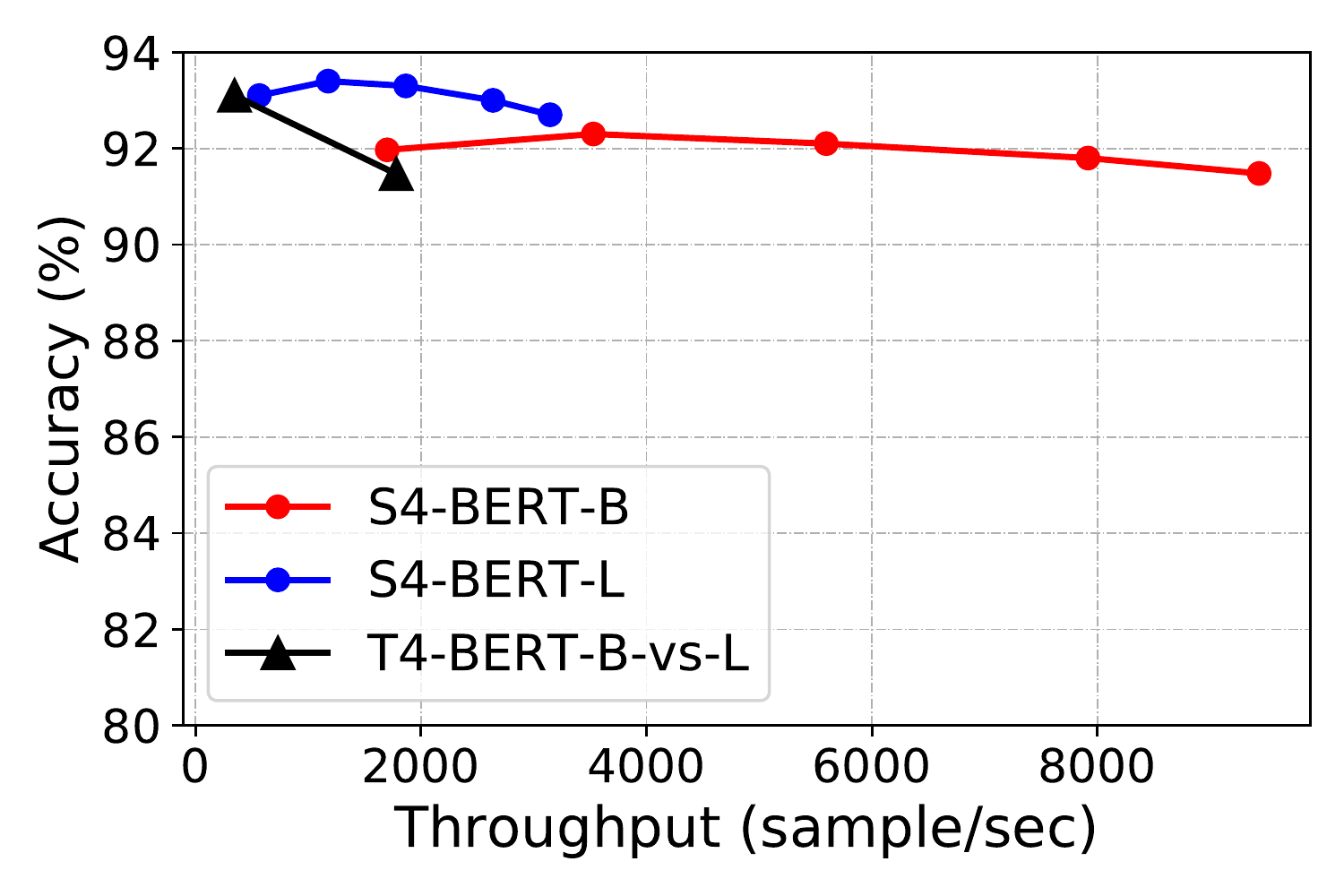}
  }
\caption{Accuracy and throughput of models of different sizes on Nvidia T4, and their sparse equivalents on Moffett S4 under sparsity=1, 2, 4, 8, 16 respectively.} 
\label{fig:acc_vs_sparsity}
\vspace{-0cm}
\end{figure*}


The most important feature of Moffett S4 is that its native support of sparse tensor representation in Tensor Cores, which only keeps the non-zero part of tensors, and therefore the degree of sparsity in a neural network directly affect the size of memory footprint, I/O cost and computation time when it is deployed on S4. Figure~\ref{fig:perf_vs_sparsity} shows the practical speedup achieved on S4 when running two benchmark models widely used in CV and NLP, respectively --- ResNet50 and BERT. Note the speedup is almost linear w.r.t. sparsity for ResNet50 and is sublinear for BERT since it has significant workloads on non-matrix-multiplication operations.

In practice, sparse model pruning achieves better accuracy-speed tradeoff than that of structured model pruning. The most common approach of reducing model size is to reduce number of layers (i.e. depth) or number of channels (i.e. width) of a neural network. For example, Figure ~\ref{fig:acc_vs_sparsity} shows accuracy and speed of Resnet50, Resnet152, BERT-base, and BERT-large, which compares the accuracy and speed achieved by \emph{dense models on T4} and \emph{sparse models on S4}. One insight from Figure ~\ref{fig:acc_vs_sparsity} is that the larger \emph{sparse} models achieve both higher accuracy and higher throughput than the smaller \emph{dense} models, which implies: \emph{a sparse model should be always considered no matter the goal is to improve accuracy or to improve speed}.
\section{Sparsification Methods}

In this section, we introduce common sparse pruning techniques that are complementary for sparse acceleration on S4. There are two scenarios that focus on different challenges of sparse pruning respectively --- (i) pruning a model trained from scratch, and (ii) pruning a model finetuned from a pretrained model. The risk of former scenario is underfiting, while the risk of latter is overfitting during pruning process.

\paragraph{Training from Scratch}

A model is trained from scratch means it is the direct solution to an optimization problem defined by the training data. Pruning such model essentially solves almost the same optimization problem but with additional \emph{sparsity constraint}, where the original dense model only plays the role of a good initialization \cite{frankle2018lottery}. Therefore, the key challenge is how to design a good optimization algorithm that fits the training data as good as the dense model under sparsity constraint. Various optimization methods have been proposed \cite{zhu2017prune,gale2019state,ma2021effective}, where sparse pruning can reduce number of parameters by an order of magnitude without significant loss of accuracy, resulting an better accuracy-efficiency tradeoff than a dense-smaller model.

\paragraph{Pretrain-Finetune Paradigm}

\begin{table*}[!t]
\addtolength{\tabcolsep}{-2pt}
\renewcommand{\arraystretch}{1.1}
\centering
\vspace{-0cm}
\begin{tabular}{lccccccc}
\toprule
\multirow{2}*{Method} & Size & MNLI-m & QNLI & MRPC  & RTE &  CoLA & Avg. \\
                     &  Reduction & (Acc) & (Acc) & (F1) & (Acc) & (Mcc) & (Acc) \\
\midrule
\multicolumn{8}{c}{\textit{\textcolor{blue}{Without Pruning}}} \\
BERT-base       & -  & 84.5  & 91.8  & 88.6  & 69.3  &  56.3  & 78.1  \\
\midrule
\multicolumn{8}{c}{\textit{\textcolor{blue}{Structural Pruning}}} \\
BERT$_6$-PKD    & 2x  & 81.5  & 89.0  & 85.0  & 65.5  &  45.5  & 73.3 \\
Theseus & 2x  &  82.3   & 89.5  & 89.0  & 68.2  &  51.1  & 76.0   \\
MiniLM$_6$      & 2x   & 84.0  & 91.0  & 88.4  & 71.5  &  49.2  & 76.8 \\
TinyBERT$_6$    & 2x   & 84.5 & 90.4  & 87.3  & 66.0  &  54.0  & 76.4  \\
TinyBERT$_4$    & 5.6x &  83.8  & 88.7  & 86.8  & 66.5  &  49.7  & 75.1   \\
\midrule
\multicolumn{8}{c}{\textit{\textcolor{blue}{Sparse Pruning}}} \\
\textbf{SparseBERT}     & \textbf{16x}  & \textbf{83.5}   & \textbf{90.8}  & \textbf{88.5}  & \textbf{69.1}  &  \textbf{54.0}  & \textbf{77.2}  \\
\bottomrule
\end{tabular}
\caption{Comparison on the dev sets of GLUE.}\vspace{0cm}
\label{tb:glue_res}
\vspace{-0cm}
\end{table*}

Pre-trained models such as BERT~\cite{devlin2019bert} and ViT~\cite{dosovitskiy2020image}, 
have become standard and effective methods for improving performance on a variety of NLP and CV tasks. These models are pre-trained in a self-supervised manner and then fine-tuned for downstream tasks. There are two approaches of pruning under the paradigm ~\cite{ganesh2021compressing}: (i) pruning during pre-training and (ii) pruning during fine-tuning on the downstream task. However, both approaches are challenged from different perspectives: pruning during pre-training suffers from underfitting since the model needs to learn not only task-related knowledge but also unrelated part during pretrain phase ~\cite{NEURIPS2020_b6af2c97}; on the other hand, pruning on downstream data suffers from overfitting as the downstream training data might not contain knowledge learned at pretraining phase ~\cite{huang2021sparse}.

State-of-the-art method typically designs pruning objectives to keep not only knowledge in downstream data but also \emph{transferred knowledge} from pretraining data. One simple method to do this is via knowledge distillation of intermediate layers ~\cite{xu2021rethinking}, which requires pruning to keep not only the prediction of data but also the intermediate feature maps generated by the pretrained model. 
We adopted the method of ~\cite{xu2021rethinking} to give pruning results on a couple of GLUE data sets in Table ~\ref{tb:glue_res} \footnote{SparseBERT adopts the training pipeline and reference results from an older version of TinyBERT \cite{jiao2019tinybertv4}.} to compare with structured distillation methods: Bert-of-Theseus~\cite{xu2020bert}, MiniLM~\cite{wang2020minilm}, and TinyBERT~\cite{jiao2020tinybert}, where sparse pruning achieves not only more reduction of model sizes but also higher prediction accuracy.
\vspace{-0.15em}

\section{Conclusion}
\vspace{-0.15em}

We introduce S4, the first commercial hardware platform to support a high degree of sparsity acceleration for deep neural network models. The S4 card is equipped with Moffett's first Antoum processor. To support high-rate sparse tensor operations while achieving high model accuracy, the S4 has a high-rate sparse tensor kernel. To have high-performance multimedia processing capabilities, the S4 card integrates dedicated hardware video codec engines and JPEG decoder engines. S4 has a scalable multi-channel subsystem that flexibly supports model parallelism and data parallelism. Combined with state-of-the-art sparse pruning technology, we demonstrate that the S4 is several times faster than mainstream inference platforms in real-world inference in CV and NLP applications.







\newpage
{
\small
\bibliography{reference}
\bibliographystyle{unsrt}
}

\end{document}